\documentclass[10pt,twocolumn,letterpaper]{article}

\usepackage{cvpr}              % To produce the CAMERA-READY version
\usepackage{multirow} 
\usepackage[dvipsnames]{xcolor}

\definecolor{cvprblue}{rgb}{0.21,0.49,0.74}
\usepackage[pagebackref,breaklinks,colorlinks,citecolor=cvprblue]{hyperref}
\usepackage{pifont}
\def\paperID{*****} % *** Enter the Paper ID here
\def\confName{CVPR}
\def\confYear{2024}

\title{DVMSR: Distillated Vision Mamba for Efficient Super-Resolution}

\author{Xiaoyan Lei{\textsuperscript{1}}, Wenlong Zhang{\textsuperscript{2}}\thanks{Corresponding author}, Weifeng Cao{\textsuperscript{1}}$^*$\\
{\textsuperscript{1}} Zhengzhou University of Light Industry,
{\textsuperscript{2}} The HongKong Polytechnic University \\
{\tt\small xyan\_lei@163.com, wenlong.zhang@connect.polyu.hk, 
weifeng\_cao@163.com}
}

\begin{document}
\maketitle
\begin{abstract}

Efficient Image Super-Resolution (SR) aims to accelerate SR network inference by minimizing computational complexity and network parameters while preserving performance. Existing state-of-the-art Efficient Image Super-Resolution methods are based on convolutional neural networks. Few attempts have been made with Mamba to harness its long-range modeling capability and efficient computational complexity, which have shown impressive performance on high-level vision tasks. In this paper, we propose DVMSR, a novel lightweight Image SR network that incorporates Vision Mamba and a distillation strategy. The network of DVMSR consists of three modules: feature extraction convolution, multiple stacked Residual State Space Blocks (RSSBs), and a reconstruction module. Specifically, the deep feature extraction module is composed of several residual state space blocks (RSSB), each of which has several Vision Mamba Moudles(ViMM) together with a residual connection. To achieve efficiency improvement while maintaining comparable performance, we employ a distillation strategy to the vision Mamba network for superior performance. Specifically, we leverage the rich representation knowledge of teacher network as additional supervision for the output of lightweight student networks. 
Extensive experiments have demonstrated that our proposed DVMSR can outperform state-of-the-art efficient SR methods in terms of model parameters while maintaining the performance of both PSNR and SSIM. The source code is available at \url{https://github.com/nathan66666/DVMSR.git}

\end{abstract}

\section{Introduction}

\begin{figure}[t]
    \centering
  \includegraphics[width=0.95\linewidth]{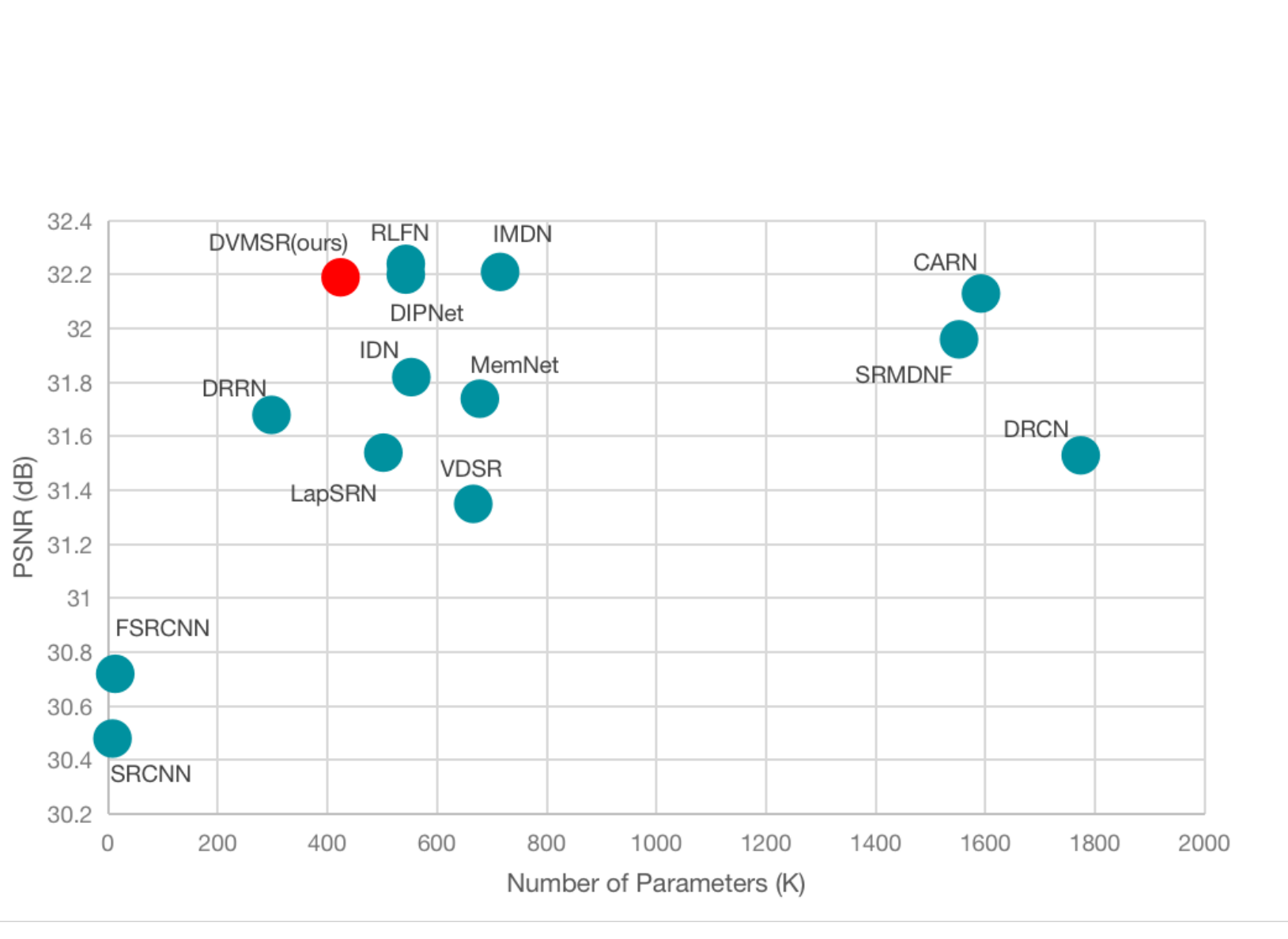}  
  \caption{PSNR results v.s the total number of parameters of different methods for image SR on Set5.}
  \label{fig:tabparam}
\end{figure}

Single image super-resolution (SR) is a key challenge in computer vision and image processing, aiming to reconstruct a high-resolution image from a low-resolution input. Effective super-resolution aims to improve the efficiency of the SR model while maintaining reconstruction performance.
Since the introduction of deep learning into super-resolution tasks~\cite{SRCNN}, many CNN-based methods have been proposed~\cite{second,FSRCNN,DRCN,VDSR,accelerating,classsr,blueprint} to improve the performance.

 A series of approaches~\cite{FSRCNN,VDSR,DRCN,LapSRN,IDN,IMDN,RFDN,RLFN,dipnet} have been proposed for building efficient models for image SR. The majority of these efficient models focus on five factors: runtime, parameters, FLOPS, activations, and depths. To further promote the development of efficient SR, ICCV holds the first competition in the AIM 2019 challenge~\cite{zhang2019aim}. The information multi-distillation network(IMDN)~\cite{IMDN} proposes cascaded information multi-distillation blocks to improve the feature extraction module, which won first place in this competition. After that, The winning solution of the AIM 2020 challenge~\cite{zhang2020aim}, residual feature distillation network(RFDN)~\cite{RFDN}, further improves the IMDN by residual learning in the main block. In the efficient SR track of NTIRE 2022 ~\cite{khan2022ntire} challenge, the winning solution, residual local feature network(RLFN)~\cite{RLFN}, removes the hierarchical distillation connection of residual feature distillation block(RFDB)~\cite{RFDN} to reduce the inference time.
In the efficient SR track of NTIRE 2022 ~\cite{dipnet} challenge, the winning solution utilizes a multi-stage lightweight training strategy that combines distillation and pruning to reduce both time consumption and model size.
 
The Transformer model, initially successful in natural language processing~\cite{transformer}, has attracted interest from the computer vision community. Its effectiveness in high-level visual tasks (e.g., image classification~\cite{trans1,pyramid,swint}) has demonstrated the potential in super-resolution~\cite{swinir,HAT}. Recently, Mamba~\cite{mamba} has demonstrated superior performance over Transformers across various sizes on large-scale real data and exhibits linear scalability with sequence length. Despite pioneering works adopting Mamba for vision tasks~\cite{ultralight, mamba,vlmamba}, it is still in its initial stages of exploring its potential (e.g., long-range modeling capability and efficiency) in low-level vision.

Different from the CNN-based and transformer-based methods, our goal is to explore the long-range modeling capability and efficiency of mamba-based methods for efficient SR. In this paper, we employ vision mamba as the basic architecture to enhance the model’s long-range modeling capability and efficiency. Our DVMSR consists of several stacked Residual State Space Blocks (RSSB), each containing several Vision Mamba Modules (ViMM). The ViMM includes a unidirectional SSM, a residual connection, and SiLU activation function. These elements work together to accelerate model convergence and enhance model accuracy and efficiency. 
As shown in Figure~\ref{fig:lam}, our method can achieve a larger perception range compared with other methods. Furthermore, we utilize a distillation strategy to enhance the model’s efficiency. We introduce a Mamba network with a larger number of parameters as the teacher network to extract knowledge for the learning of the student network. Extensive experiments and ablation studies have shown the effectiveness of our proposed method.

Our contributions can be summarized as follows:
\begin{enumerate}
\item By leveraging the long-range modeling capability of Vision Mamba, we propose a lightweight model with unidirectional state space models (SSM) for efficient super-resolution.

\item We propose a special feature distillation strategy to enhance the efficiency ability of vision mamba for efficient super-resolution.

\item Extensive experiments have shown that our proposed method outperforms existing state-of-the-art (SOTA) methods in terms of parameters while maintaining comparable PSNR and SSIM performance.

\end{enumerate}

\section{Related Work}
\subsection{Lightweight Super Resolution}

SRCNN~\cite{SRCNN} marks the inaugural application of deep learning algorithms in the Single Image Super-Resolution (SISR)~\cite{HAT,chen2023hat}. A series of works have been explored to apply the SR method in real scenarios, such as GAN-based SR~\cite{srgan,esrgan,zhang2019ranksrgan,zhang2021ranksrgan}, degradation model~\cite{closerlookBSR,realesrgan,bsrgan}, multi-task learning~\cite{tgsr} and systematic evaluation~\cite{zhang2023seal}. In real-world SR model deployments, the computing power of the deployed devices is often limited, such as edge devices, etc. In this case, the efficiency of the SR network becomes an important aspect. Efficient Image Super-Resolution aims to reduce the computational effort and parameters of the SR network while achieving faster inference times and maintaining high performance. FSRCNN~\cite{FSRCNN} reduces unnecessary computational costs by utilizing the deconvolution layer as the upsampling layer. VDSR~\cite{VDSR} is introduced to further improve super-resolution (SR) performance. DRCN~\cite{DRCN} achieves parameter reduction through deep recursive convolutional networks. LapSRN~\cite{LapSRN} employs a Laplacian pyramid super-resolution block for HR image reconstruction. DRRN~\cite{DRRN} employs recursive and residual network architectures, surpassing DRCN in both performance and parameter reduction. MemNet~\cite{MemNet} introduces a memory block to explicitly model long-term dependencies in CNN-based SR models. IDN~\cite{IDN} explicitly divides the preceding extracted features into two parts. IMDN~\cite{IMDN} introduces a lightweight Information Multi-Distillation Network by constructing cascaded Information Multi-Distillation Blocks. RFDN~\cite{RFDN} proposes the residual feature distillation network. RLFN~\cite{RLFN} improves its speed by eliminating hierarchical distillation connections. DIPNet~\cite{dipnet} introduces the Reparameterization Residual Feature Block, which explores the potential of complex structures during optimization while maintaining computational efficiency. Besides, they achieve first place in the NTIRE 2023 Efficient Super-Resolution Challenge~\cite{2023ntire}.

\subsection{State space models in Vision}

Recent researches have led to a surge of interest in the state space model (SSM), which has its origins in the classic Kalman filter model~\cite{kalman}. The linear scalability of State Space Models (SSMs) in handling long-range dependencies, exemplified by the Mamba architecture~\cite{mamba}, contrasts with Transformers. While Mamba outperforms Transformers in natural language tasks, recent research endeavors extend its applicability to vision tasks. Specifically, Mamba models are designed to capture long-range temporal dependencies in video data, enhancing video classification performance~\cite{videossm1,videossm2,videossm3,videossm4}. Additionally, other works explore Mamba’s applicability in vision tasks, including image classification~\cite{vmamba,vim}, biomedical image segmentation~\cite{umamba}, remote sensing image classification~\cite{rsmamba}, and Multimodal Learning~\cite{vlmamba}. The research conducted by~\cite{mambair} emphasizes Mamba’s utility as a straightforward and efficient baseline for image restoration in low-level vision tasks. Our work extends this by proposing a novel network architecture that combines Mamba with distillation, achieving a trade-off between super-resolution quality and computational efficiency.

\begin{figure*}[h]
\centering
\includegraphics[width=6in]{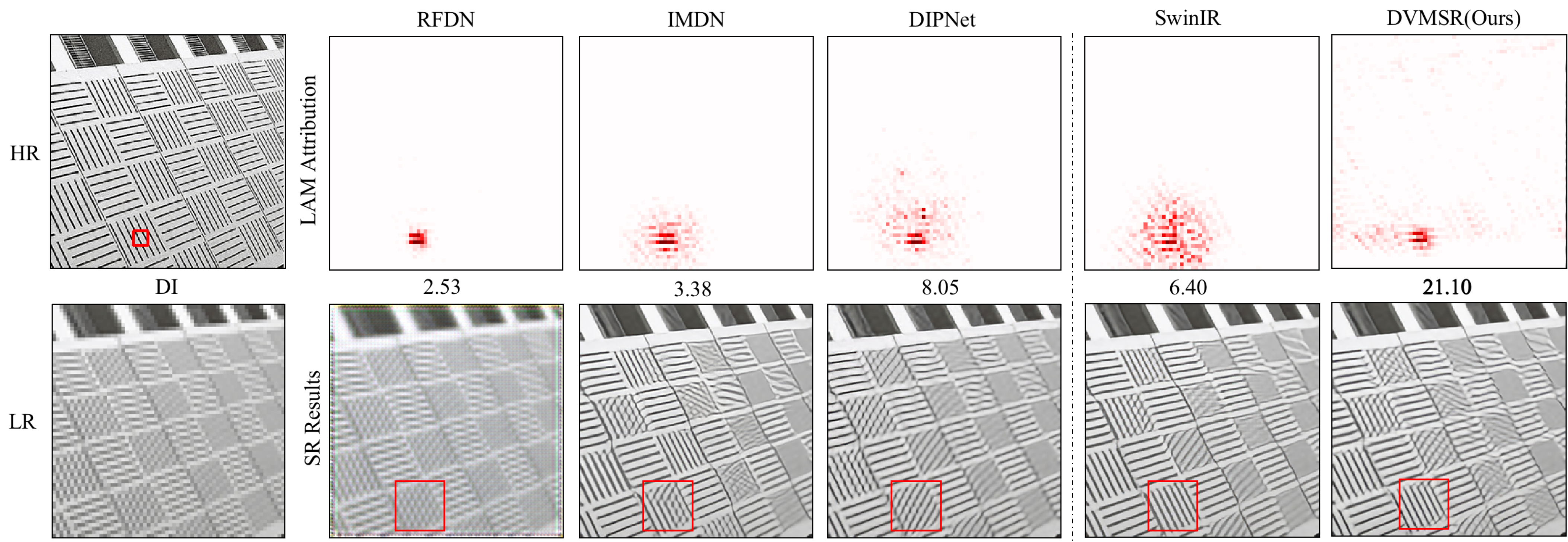}
\caption{The LAM results are provided for various networks including both CNN-based and transformer-based methods. LAM attribution indicates the significance of each pixel in the input LR image during the reconstruction process of the patch highlighted by a box. The Diffusion Index (DI) denotes the extent of pixel involvement. A higher DI indicates a broader range of utilized pixels.}
\label{fig:lam}
\end{figure*}

\subsection{Feature Distillation}
Knowledge distillation stands out as a straightforward yet powerful technique for enhancing the performance of smaller models, a necessity driven by the limited computing power of deployed devices. This method involves training a smaller network (student) under the guidance of a larger network (teacher), enabling effective knowledge transfer. Unlike other compression methods, knowledge distillation can reduce network size regardless of structural differences between the teacher and student networks. The seminal work by \cite{distilling} introduced the knowledge distillation (KD) method, utilizing the softmax output of the teacher network. Notably, this method can be applied across various network architectures due to matching output dimensions. Over time, intermediate layer distillation methods have emerged, leveraging insights from the teacher network's convolutional or penultimate layers, preserving crucial feature-map localities~\cite{para,paying,fitnets,distilling}. Moreover, there exists a wealth of research integrating distillation techniques into super-resolution tasks~\cite{srdis1,srdis2,srdis3,srdis4,srdis5}. In this paper, we focus on adopting the output feature map of a pre-trained model as the distillation target. Through extensive experimentation, we demonstrate the effectiveness of our approach in enhancing model performance.

\section{Methodology}

\subsection{Motivation}

Efficient Super Resolution (SR) is designed to transform low-quality images into high-quality counterparts, leveraging a small parameter set and minimal computational power. ESR predominantly relies on CNNs for local feature extraction, but their limited long-range modeling hinders performance. Transformers, while proficient in global context, introduce computational complexities.

Mamba excels in high-level vision tasks, supported by prior research~\cite{vmamba,vim,umamba,ultralight,rsmamba,vlmamba}. 
Motivated by Mamba’s long-range modeling capabilities, we investigate its performance in super-resolution (SR) tasks, comparing it to CNN-based ESR methods~\cite{RFDN, IMDN, dipnet} and transformer-based method~\cite{swinir}. To elucidate Mamba's operational mechanisms, we employe a specialized diagnostic tool called LAM~\cite{lam}, designed specifically for SR tasks. Utilizing LAM enabled us to pinpoint the input pixels that contribute most significantly to the selected region. As depicted in Figure~\ref{fig:lam}, the red-marked points denote informative pixels crucial for the reconstruction process. Notably, DVMSR exhibited a notably higher DI (Diffusion index) indication compared to other models, indicating its superior ability to leverage a broader range of pixel information and affirming its exceptional long-range modeling capability. The proposed DVMSR yields improved image details during the reconstruction process, thereby substantiating its efficacy for super-resolution tasks.

\begin{figure*}[!htbp]
\centering
\includegraphics[width=6.5in]{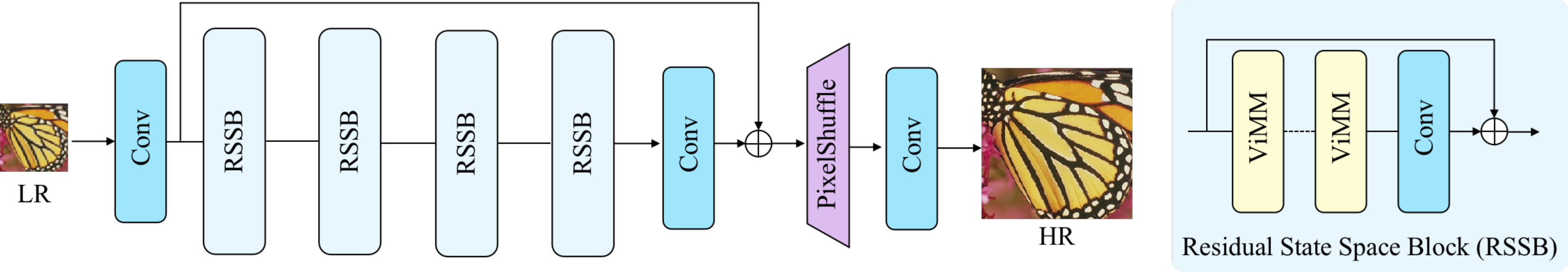}
\caption{The overall network architecture of our DVMSR.}
\label{fig:dvmsr}
\end{figure*}

\subsection{Preliminaries}
State space models (SSMs), such as the Mamba deep learning model, hold potential for long sequence modeling. Inspired by continuous systems, SSMs map a 1-D function or sequence $x(t)\in \mathbb{R} \longmapsto y(t)\in \mathbb{R}$ via a hidden state $h(t)\in \mathbb{R}^N $. The formulation is as follows:

\begin{equation} \label{eq1}
  \begin{split}
  & h'(t)=Ah(t)+Bx(t), \\
  & y(t)=Ch(t).
  \end{split}
\end{equation}
where N is the state size, $A\in \mathbb{R}^{N\times N} $, $B\in \mathbb{R}^{N\times 1} $, $C\in \mathbb{R}^{1\times N} $. 

Mamba is the discrete versions of the continuous system, and it achieves this by utilizing $\Delta$  to convert continuous parameters A and B into their discrete counterparts, $\bar{A}$ and $\bar{B}$.  The commonly used method for transformation is zero-order hold (ZOH), which is defined as follows:

\begin{equation} \label{eq2}
  \begin{split}
  & \bar{A} =exp(\Delta A),\\
  & \bar{B}=(\Delta A)^{-1}(exp(\Delta A)-I)\cdot \Delta B.
  \end{split}
\end{equation}

After the discretization of $\bar{A}$, $\bar{B}$, the discretized version of Eq.~\ref{eq1} using a step size $\Delta$ can be rewritten as:

\begin{equation} \label{eq3}
  \begin{split}
  & h_t=\bar{A}h_{t-1}+\bar{B}x_t, \\
  & y_t=Ch_t.
  \end{split}
\end{equation}

\subsection{Overall network architecture}

The overall network architecture of our proposed DVMSR is depicted in Figure~\ref{fig:dvmsr}. Our DVMSR mainly consists of three main modules: feature extraction convolution, multiple stacked Residual State Space Blocks (RSSBs), and a reconstruction module. Specifically, for a given low-resolution (LR) input $I_{LR}\in \mathbb{R}^{H\times W\times C_{in}}$ , we exploit one convolution layer to extract the first feature $F_{0} \in \mathbb{R}^{H\times W\times C}$, where $C_{in }$ and $C$ denote the channel number of the input and the intermediate feature. Then, a series of Residual State Space Block (RSSB) and one $3\times 3$ convolution layer $HConv(\cdot)$ are utilized to perform the deep feature extraction. After that, we add a global residual connection to fuse shallow features $F_{0}$ and deep features $F_{D}\in \mathbb{R}^{H\times W\times C}$, and then reconstruct the high-resolution result via a reconstruction module. As depicted in Figure~\ref{fig:dvmsr}, each RSSB contains two Vision Mamba Module (ViMM) and a $3\times 3$ convolution layer with a residual connection. For the reconstruction module, the pixel-shuffle method is adopted to up-sample the fused feature.

\begin{figure}[!htbp]
\centering
\includegraphics[width=2.0in]{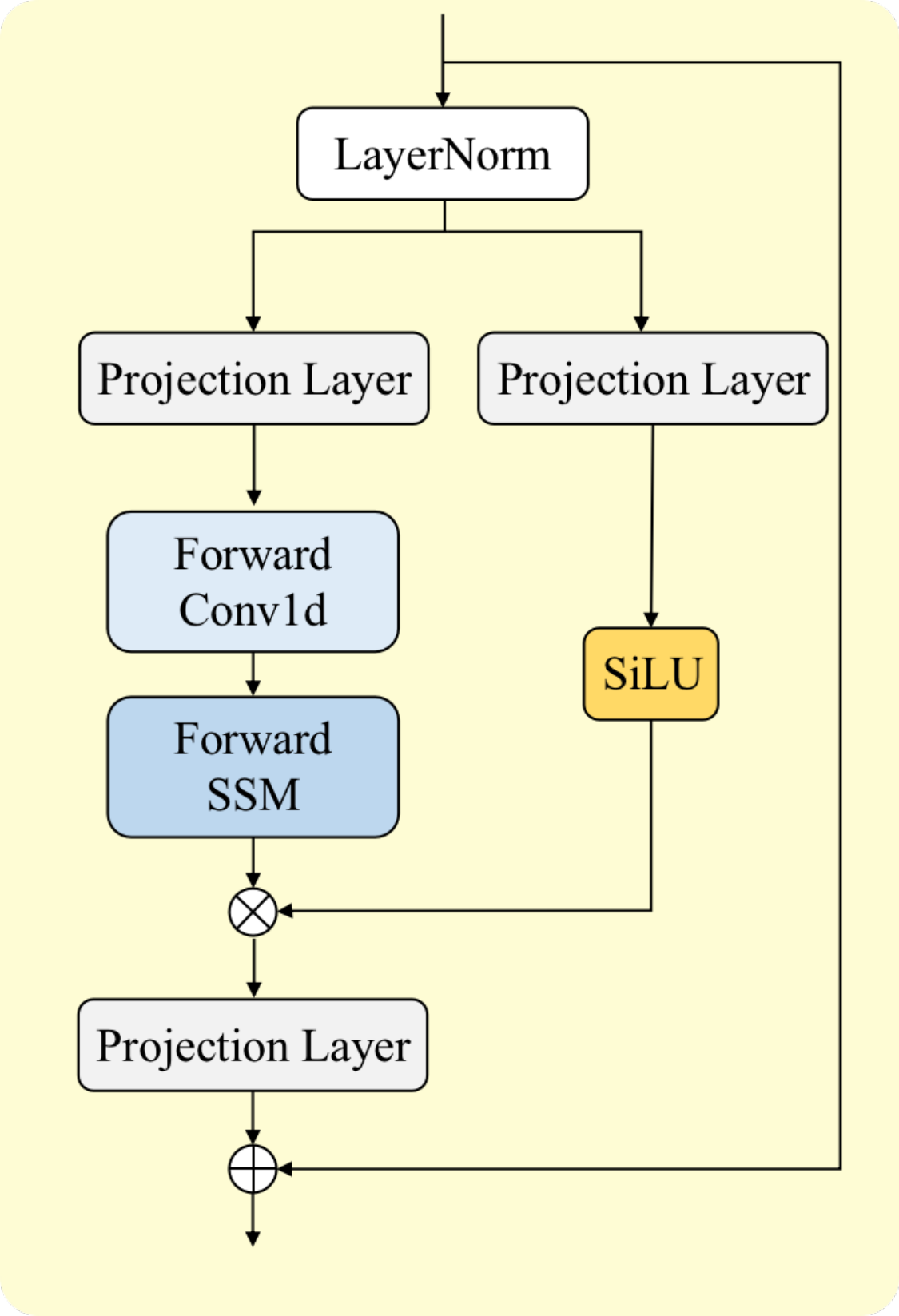}
\caption{The structure of Vision Mamba Module(ViMM).}
\label{fig:vimm}
\end{figure}
\subsubsection{Mamba network} The design of mamba network is shown in Figure~\ref{fig:vimm}, which is Vision Mamba Module (ViMM) using unidirectional sequence modeling. The input token sequence $X \in \mathbb{R}^{H\times W\times C}$ is first normalized by the normalization layer. Next, we linearly project the normalized sequence, expanded the features channel to $\lambda C$. We proceed by processing the projection layer through 1-D convolution, resulting in the computation of $X_1$ via the SSM. The $X_1$ gated by the projection layer and a residual connection to get the output token sequence $X_{out}\in \mathbb{R}^{H\times W\times C}$, as follows:

\begin{equation} \label{eq5}
  \begin{split}
   & X_1 = SSM(Conv1d(Linear(LN(X)))),\\
   & X_2 = SiLU(Linear(LN(X))),\\
   & X_{out} = Linear(X_1 \odot X_2) + X.
  \end{split}
\end{equation}
Where $LN$ is the layer normalization and $\odot$ denotes the Hadamard product.

\begin{figure*}[!htbp]
\centering
\includegraphics[width=6.5in]{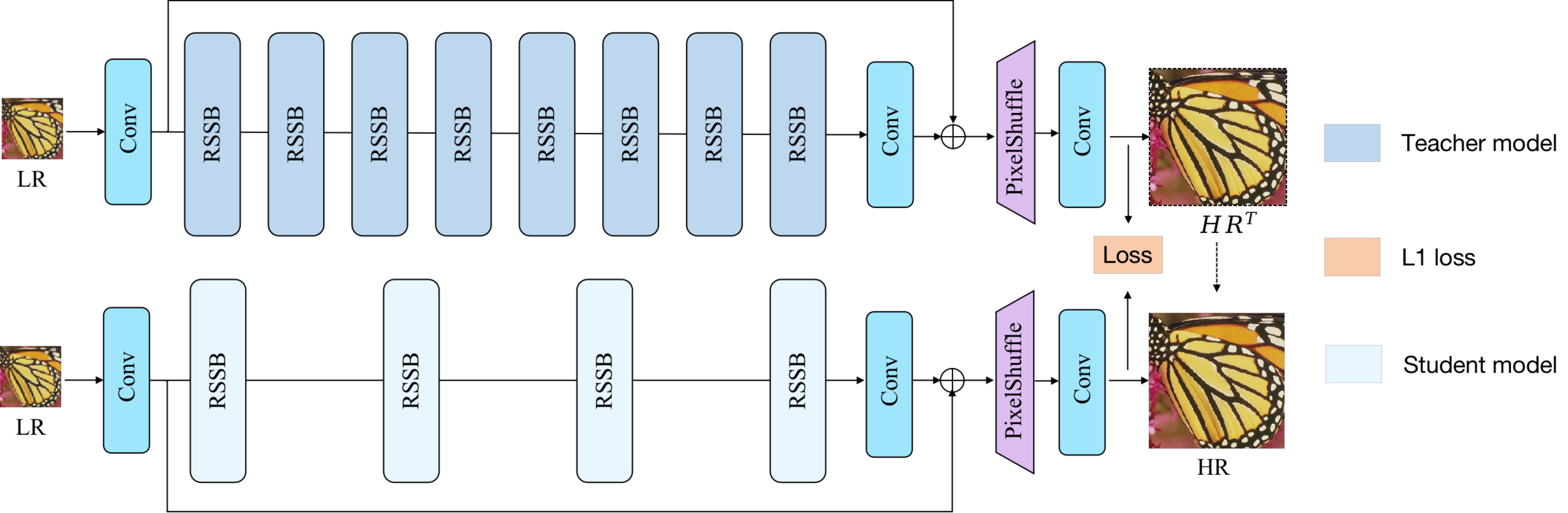}
\caption{The deep feature distillation pipeline of our method.}
\label{fig:tsmodel}
\end{figure*}

\subsubsection{Distillation strategy}
Our method introduces a deep feature distillation strategy (Fig.~\ref{fig:tsmodel}). During the distillation stage, the teacher network accumulates rich representation knowledge, maintaining a fixed state. By minimizing the $\mathcal{L}_{1}$ loss, we ensure alignment between student network features and those of the teacher. This formal process facilitates effective knowledge transfer from the teacher to the student network: 

\begin{equation} \label{eq1}
  \begin{split}
  & \mathcal{L}_{out}= \lambda _{dis} \mathcal{L}_{dis}+\lambda _{1}\mathcal{L}_1, \\
  & \mathcal{L}_{dis} = \left \|  \mathcal{T}(I_{LR})-\mathcal{S}(I_{LR})\right \|_{1},\\
  & \mathcal{L}_1= \left \| I_{HR}-\mathcal{S}(I_{LR})\right \|_{1},\\
  \end{split}
\end{equation}
where $\lambda _{dis}$ and $\lambda _{1}$ represents the coefficient of the $\mathcal{L}_{dis}$ loss function and the coefficient of the $\mathcal{L}_{1}$ loss function, respectively. They are set 1. $\mathcal{T}$ represents the function of our teacher network and $\mathcal{S}$ denotes the function of our proposed network. $I_{LR}$ and $I_{HR}$ are the input LR images and the corresponding ground-truth HR images, respectively. More information of $\mathcal{L}_{dis}$ can be seen from Fig.\ref{fig:distillation}.

\section{Experiments}

\begin{table*}[t]
    \centering
    \caption{Average PSNR/SSIM for scale factor 4 on datasets Set5, Set14, BSD100, Urban100, and Manga109. The best and second best results are highlighted in {\color{red}red} and {\color{blue}blue} respectively.}
    % The best and second best results are highlighted in {\color{red}red} and {\color{blue}blue} respectively.}
    \resizebox{0.9\linewidth}{!}{
    \begin{tabular}{lccccccc}
    \hline
    \multirow{2}{*}{Method}  & \multirow{2}{*}{Params} & Set5 & Set14 & BSD100 & Urban100 & Manga109 \\
    % \cline{4-8} 
    & \multicolumn{1}{l}{}   & PSNR/SSIM & PSNR/SSIM & PSNR/SSIM & PSNR/SSIM & PSNR/SSIM \\
    % \hline
    \hline
    \hline
    Bicubic  & - & 28.42/0.8104 & 26.00/0.7027 & 25.96/0.6675 & 23.14/0.6577 & 24.89/0.7866 \\
    SRCNN~\cite{SRCNN}  & 8K & 30.48/0.8626 & 27.50/0.7513 & 26.90/0.7101 & 24.52/0.7221 & 27.58/0.8555 \\
    FSRCNN~\cite{FSRCNN}  & 13K & 30.72/0.8660 & 27.61/0.7550 & 26.98/0.7150 & 24.62/0.7280 & 27.90/0.8610 \\
    VDSR~\cite{VDSR}  & 666K & 31.35/0.8838 & 28.01/0.7674 & 27.29/0.7251 & 25.18/0.7524 & 28.83/0.8870 \\
    DRCN~\cite{DRCN}  & 1774K & 31.53/0.8854 & 28.02/0.7670 & 27.23/0.7233 & 25.14/0.7510 & 28.93/0.8854 \\
    LapSRN~\cite{LapSRN}  & 502K & 31.54/0.8852 & 28.09/0.7700 & 27.32/0.7275 & 25.21/0.7562 & 29.09/0.8900 \\
    DRRN~\cite{DRRN}  & 298K & 31.68/0.8888 & 28.21/0.7720 & 27.38/0.7284 & 25.44/0.7638 & 29.45/0.8946 \\
    MemNet~\cite{MemNet}  & 678K & 31.74/0.8893 & 28.26/0.7723 & 27.40/0.7281 & 25.50/0.7630 & 29.42/0.8942 \\
    IDN~\cite{IDN}  & 553K & 31.82/0.8903 & 28.25/0.7730 & 27.41/0.7297 & 25.41/0.7632 & 29.41/0.8942 \\
    %EDSR &  & 1518K & 32.09/0.8938 & 28.58/0.7813 & 27.57/0.7357 & 26.04/0.7849 & 30.35/0.9067 \\
    SRMDNF~\cite{SRMDNF}  & 1552K & 31.96/0.8925 & 28.35/0.7787 & 27.49/0.7337 & 25.68/0.7731 & 30.09/0.9024 \\
    CARN~\cite{CARN}  & 1592K & 32.13/0.8937 & 28.60/0.7806 & {27.58}/0.7349 & 26.07/0.7837 & 30.47/{0.9084} \\
    % SwinIR~\cite{swinir}  & 897K & 32.44/0.8976 & 28.77/0.7858 & 27.69/0.7406 & 26.47/0.7980 & 30.92/0.9151 \\
    IMDN~\cite{IMDN}  & 715K & {\color{blue}32.21}/{0.8948} & 28.58/0.7811 & 27.56/0.7353 & 26.04/0.7838 & 30.45/0.9075 \\
    RFDN~\cite{RFDN} & 550K & 32.24/0.8952 & 28.61/0.7819 & 27.57/0.7360 & 26.11/0.7858 & {\color{red}30.58}/{\color{red}0.9089} \\
    RLFN~\cite{RLFN}  & 543K & {\color{red}32.24}/{\color{blue}0.8952} & {\color{red}28.62}/0.7813 & {\color{red}27.60}/0.7364 & {\color{red}26.17}/{\color{blue}0.7877} & -/- \\
    DIPNet~\cite{dipnet}  & 543K & 32.20/0.8950& 28.58/0.7811 &{\color{blue}27.59}/0.7364 &{\color{blue}26.16}/{\color{red}0.7879}& {\color{blue}30.53}/{\color{blue}0.9087}\\
    
    % DVMSR (Ours)  & 424K & 32.16/0.8955 & 28.60/0.7821 & 27.57/0.7377 & 26.01/0.7834 &  30.45/0.9079 \\
    DVMSR (Ours)  & 424K & 32.19/{\color{red}0.8955} & {\color{blue}28.61}/{\color{red}0.7823} & 27.58/{\color{red}0.7379} & 26.03/0.7838 &  30.48/0.9084 \\
    \hline
    \end{tabular}}
    \label{tab:sota}
\end{table*}

\subsection{Datasets and metrics}
In this paper, DF2K (DIV2K + Flickr2K)~\cite{div2k} with 3450 images are used for training the proposed model from scratch. During testing, we select five standard benchmark datasets: Set5~\cite{Set5}, Set14~\cite{Set14}, BSD100~\cite{B100}, Urban100~\cite{Urban100} and Manga109~\cite{Manga109}. The low-resolution images are generated from the ground truth images by the ``bicubic'' downsampling in MATLAB. PSNR/SSIM measured by discarding a 4-pixel boundary around the images, and calculated on the Y channel is reported for the quantitative metrics.
 
\subsection{Implementation details}
During training, we set the input patch size to 256 $\times$ 256 and use random rotation and horizontal flipping for data augmentation. The batch size is set to 128 and the total number of iterations is 500k. The initial learning rate is set to $2\times 10^{-4}$. We adopt a multi-step learning rate strategy, where the learning rate will be halved when the iteration reaches 250000, 400000, 450000, and 475000, respectively. Adam optimizer with $\beta _{1} = 0.9$ and $\beta _{2} = 0.99$ is used to train the model. 

\noindent\textbf{Distillation training.} In the teacher learning phase, we utilize the DF2K dataset with 2K resolution to train the teacher network, which comprises 8 RSSB and 2 ViMM blocks with 192 channels. During the distillation training phase, we use DF2K datasets for the student network, which contains 4 RSSB and 2 ViMM blocks with 60 channels.

\subsection{Comparison with State-of-the-art SR models}

We compare DVMSR with several advanced efficient super-resolution model~\cite{SRCNN,FSRCNN,VDSR,DRCN,LapSRN,DRRN,MemNet,IDN,SRMDNF,CARN,IMDN,RFDN,RLFN,dipnet}. The quantitative performance comparison on several benchmark datasets~\cite{Set5,Set14,B100,Urban100,Manga109} is indicated in Table~\ref{tab:sota}. Our experimental results showcase our ability to achieve smaller parameter counts while surpassing several previous methods on five benchmark datasets. Specifically, we attained higher SSIM scores on Set5, Set14, and BSD100. It's important to note that SSIM scores serve as a crucial metric, indicating how effectively our model preserves the structure and content of the images, ultimately resulting in reconstructions that closely resemble the original images. Additionally, we observed that PSNR values remain comparable across these five datasets. This comprehensive evaluation underscores the effectiveness of our approach in enhancing image quality while maintaining efficiency, making it a promising solution for various image enhancement tasks. It's worth emphasizing that in our current study, we directly utilize the final model architecture employed in the NTIRE competition. Remarkably, we manage to maintain excellent performance without unnecessarily inflating the parameter count. This strategic decision underscores our commitment to efficiency and effectiveness in model design, ensuring that our approach remains practical and scalable for real-world applications.

\noindent\textbf{Model complexity comparisons between SwinIR and DVMSR.} Our investigation focuses on Mamba’s performance in super-resolution (SR) tasks. In Fig.~\ref{fig:lam}, we show the excellent long-range modeling capabilities of our DVMSR using LAM. Additionally, we compare DVMSR with SwinIR, a transformer-based model, in terms of model complexity. SwinIR outperforms DVMSR by 0.23 dB in PSNR, but at the cost of approximately twice the number of parameters, significantly higher FLOPS, and about 20 times longer inference time. These findings suggest that Mamba-based models hold promise for efficient SR.

% Model complexity comparisons between transformer-based model and Mamba-based(ours).
\begin{table}[h]
    \centering
    \caption{Model complexity comparisons between SwinIR and DVMSR. Times represent the average inference time measured on the DIV2K dataset with an Nvidia RTX 3090 in seconds (s). FLOPS and memory is measured when the input is $256\times 256$. PSNR is the result of testing on DIV2K.}
     \resizebox{1\linewidth}{!}{
    \begin{tabular}{lcccccc}
        \toprule
        Method & PSNR & Time (s) & Params[M] & FLOPS[G] & Activations & Memory[M] \\
        \midrule
        SwinIR &  29.20 dB & 0.865 & 0.9296 &  70.7828& 26.7387 & 1454.458  \\ %set5 32.04/0.8940  set14 28.51/0.7801
        DVMSR & 28.97 dB & 0.048  & 0.4244  & 20.1680 & 26.7387 & 1094.245 \\   %set5 32.11/0.8944  set14 28.51/0.7805
        \bottomrule
    \end{tabular}}
    \label{tab:swinir}
\end{table}

\subsection{Ablation Study}
\subsubsection{Model Parameter Analysis}

Here, we train DVMSR on DIV2K for classical image SR (×4) and test it on Set5 and Set14.

\noindent\textbf{Impact of ViMM number.}
We show the effects of ViMM number in each RSSB on model performance in Table~\ref{tab:vimm}. In experiments 1 - 3, it is observed that the PSNR/SSIM is negatively correlated with the number of ViMMs. However, when we set the ViMM number to 1, as presented in experiment 4, the PSNR in Set5 and Set14 decreased by 0.09 dB compared to when the ViMM number is set to 2. Therefore, there may be a balance point for the ViMM number, where it should not be too large to avoid over-complexity of the model, nor too small to limit the model's ability to represent the data. Experimental results indicate that setting the ViMM number to 2 is appropriate.
\begin{table}[h]
    \centering
    \caption{Impact of ViMM number in each RSSB on the Set5 and Set14 datasets with scale factor of $\times 4$. The number of RSSB is fixed at 4 and keep other parameter settings consistent. The best results are highlighted.}
    \resizebox{0.95\linewidth}{!}{
    \begin{tabular}{lcccccc}
        \toprule
        
        \multirow{2}{*}{Exp.} & \multirow{2}{*}{Params[M]} & \multirow{2}{*}{ViMM number} & Set5 & Set14\\
         & & & PSNR/SSIM & PSNR/SSIM \\
        \midrule
        1 & 7.222 & 6,6,6,6 & 31.99/0.8926 & 28.44/0.7785\\ 
        2 & 5.214 & 2,2,9,2 & 32.17/0.8959 & 28.63/0.7834 \\
        3 & 3.651 & 2,2,2,2 & \textbf{32.30/0.8972} & \textbf{28.68/0.7847} \\ 
        4 & 2.758 & 1,1,1,1 & 32.21/0.8954 & 28.59/0.7821 \\
        \bottomrule
    \end{tabular}}
    \label{tab:vimm}
\end{table}

\noindent\textbf{Impact of RSSB number.}
In Table~\ref{tab:rssb}, In Experiments 1-3, as the RSSB number increases, the parameter count increases, with the channel number set to 180. Along with the increase in RSSB number, the PSNR in Set5 shows a significant improvement. Compared to Experiment 1, Experiment 2 shows an increase of 0.26 dB, and relative to Experiment 2, Experiment 3 shows an increase of 0.13 dB. When we set the RSSB number to 10, the improvement is moderated, with Experiment 4 showing an increase of 0.01 dB relative to Experiment 3.

\begin{table}[h]
    \centering
    \caption{Impact of RSSB number on the Set5 and Set14 datasets with scale factor of $\times 4$. The number of ViMM is fixed at 2 and keeps other parameter settings consistent. The best results are highlighted.}
    \resizebox{0.9\linewidth}{!}{
    \begin{tabular}{lcccccc}
        \toprule
        \multirow{2}{*}{Exp.} & \multirow{2}{*}{Params[M]} & \multirow{2}{*}{RSSB number} & Set5 & Set14\\
         & & & PSNR/SSIM & PSNR/SSIM \\
        \midrule
        1 & 2.175 & 2 & 32.04/0.8938 & 28.51/0.7799 \\ 
        2 & 3.651 & 4 & 32.30/0.8972 & 28.68/0.7847 \\
        3 & 5.128 & 6 & 32.43/0.8987 & 28.75/0.7866 \\ 
        4 & 8.080 & 10 & \textbf{32.44/0.8990} & \textbf{28.77/0.7874} \\
        \bottomrule
    \end{tabular}}
    \label{tab:rssb}
\end{table}

\noindent\textbf{Impact of channel number.}
We maintained the ViMM number and RSSB number while examining the influence of channel numbers on model performance, as detailed in Table~\ref{tab:channel}. Notably, our analysis revealed a diminishing improvement in model performance when the channel number was set to 210. Thus, we conclude that setting the channel number to 192 is more suitable for optimal model performance.

\begin{table}[h]
    \centering
    \caption{Impact of channel number on the Set5 and Set14 datasets with scale factor of $\times 4$. keep other parameter settings consistent. The best results are highlighted.}
    \resizebox{0.95\linewidth}{!}{
    \begin{tabular}{lcccccc}
        \toprule
        \multirow{2}{*}{Exp.} & \multirow{2}{*}{Params[M]} & \multirow{2}{*}{channel number} & Set5 & Set14\\
         & & & PSNR/SSIM & PSNR/SSIM \\
        \midrule
        1 & 2.664 & 150 & 32.32/0.8971 & 28.65/0.7838 \\ 
        2 & 3.651 & 180 & 32.30/0.8972 & 28.68/0.7847 \\
        3 & 4.089 & 192 & 32.37/0.8977 & \textbf{28.71/0.7851} \\ 
        4 & 4.809 & 210 & \textbf{32.39/0.8976} & 28.71/0.7850 \\
        \bottomrule
    \end{tabular}}
    \label{tab:channel}
\end{table}

\begin{table}[h]
    \centering
    \caption{ Comparison of unidirectional SSM or bidirectional SSM.  Times represent the average inference time measured on the DIV2K dataset with an Nvidia RTX 3090 in seconds (s). FLOPS and memory are measured when the input is $256\times 256$. PSNR is the result of testing on DIV2K.}
    \resizebox{1\linewidth}{!}{
    \begin{tabular}{lcccccc}
        \toprule
        Method & PSNR & Time (s) & Params[M] & FLOPS[G] & Activations & Memory[M] \\
        \midrule
        unidirectional SSM & 28.87 dB & 0.048 & 0.4244 & 20.1680 & 26.7387 & 1094.245  \\ 
        bidirectional SSM & 28.88 dB & 0.087 & 0.4849 & 23.9429 & 26.7387 & 1451.680 \\   
        \bottomrule
    \end{tabular}}
    \label{tab:bissm}
\end{table}

\subsubsection{Distillation Learning}
\textbf{Distillation loss.} To investigate the effectiveness of distillation loss, we tried multiple distillation strategies. Mid-level feature distillation and end-level feature distillation are presented in Figure~\ref{fig:distillation}. As shown in Table~\ref{tab:distillation}, using the end-level feature distillation method tends to increase the PSNR and SSIM on Set5 and Set14 datasets. This suggests that \textbf{the features towards the end of the model might be closer to the target output of the SR task.} When attempting to alter the weights and types of distillation loss in the mid-level feature distillation method, there were no changes observed in PSNR and SSIM values on Set5 and Set14 datasets. This indicates that it is difficult for the student model to benefit from the features of the middle layer of the teacher model, as even after modifying the weights and types of distillation loss, there were no significant changes in the results. When we increase the weight of distillation loss in the end-level feature distillation method, there is a slight decrease in the PSNR and SSIM on Set5 and Set14 datasets. This could be \textbf{because excessively high weights on distillation loss might introduce too many constraints, thereby affecting the model's performance.}

\begin{table}[t]
    \centering
	\begin{minipage}{0.45\linewidth}
		\centering
        \includegraphics[width=0.8\linewidth]{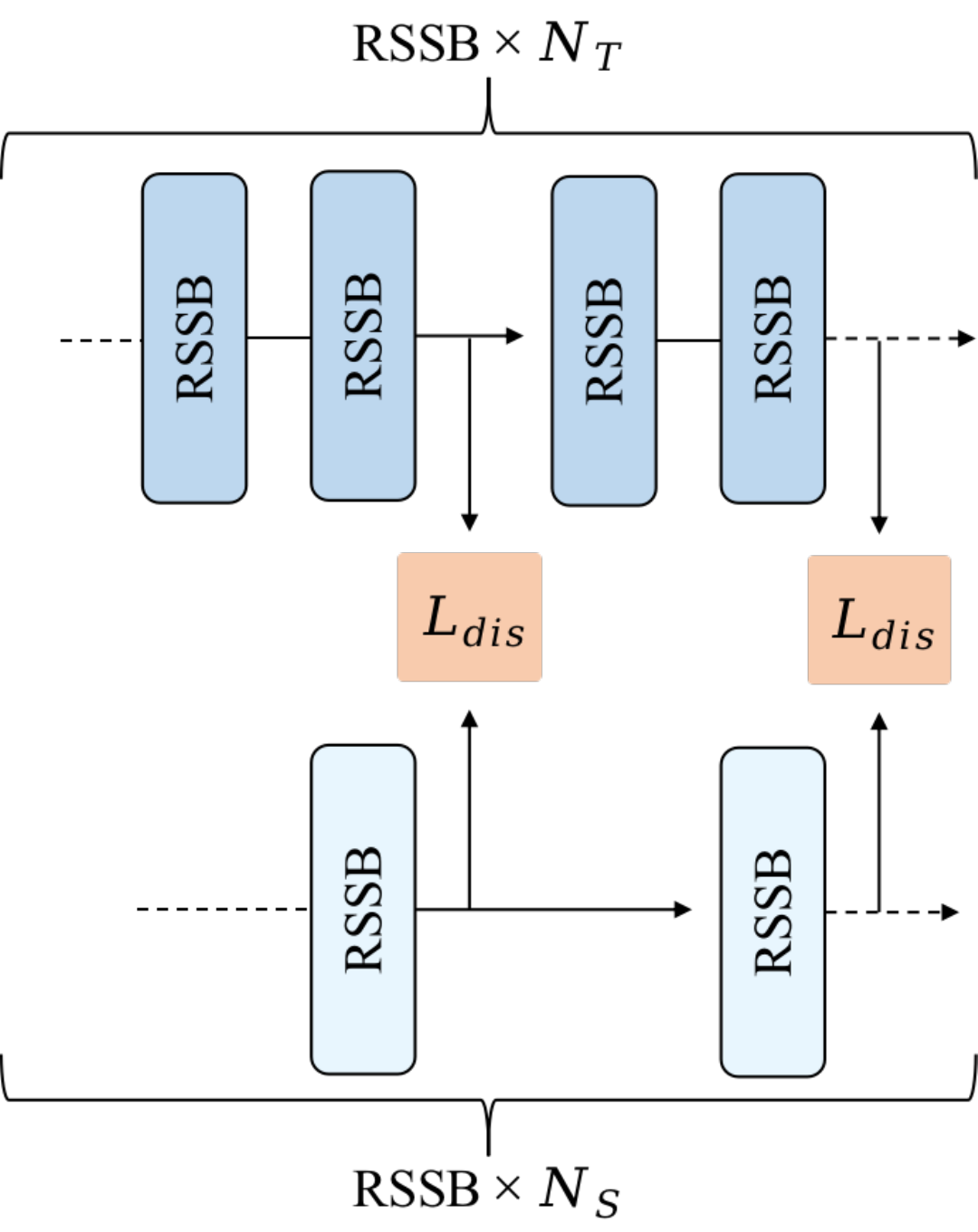}  
		% \captionof{figure}{(a)}
		\label{fig:mid_feature}
	\end{minipage}
	\begin{minipage}{0.5\linewidth}
		\centering
        \includegraphics[width=0.82\linewidth]{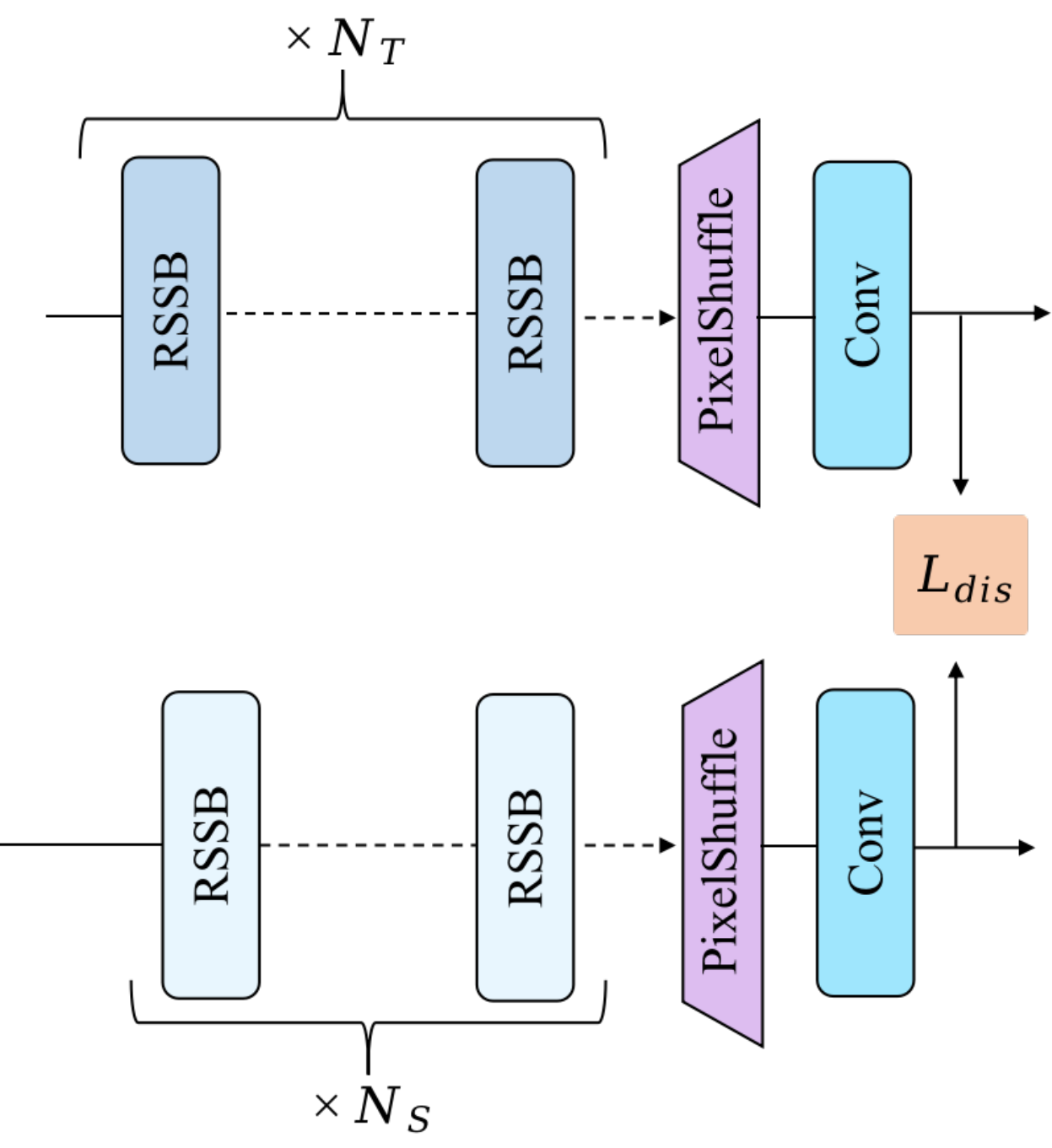}  
		% \captionof{figure}{(b)}
		\label{fig:fin_feature}
	\end{minipage}
        \captionof{figure}{Left: The structure of mid-level feature distillation; Right: The structure of end-level feature distillation}
\label{fig:distillation}
\end{table}

\begin{table}[h]
    \centering
    \caption{Impact of the distillation loss. ``\ding{56}'' signifies that distillation is not used, and ``\ding{52}'' signifies that distillation is used. ``mid'' and ``end'' represent mid-level feature distillation and end-level feature distillation, respectively. $\mathcal{L}_{dis}: \mathcal{L}_{1}$ represents the weight ratio of the distillation loss and $\mathcal{L}_{1}$ loss.}
    \resizebox{1.0\linewidth}{!}{
    \begin{tabular}{lcccccc}
        \toprule
        distillation & distillation & distillation & \multirow{2}{*}{$\mathcal{L} _{dis}:\mathcal{L} _{1}$} & Set5 & Set14\\
         strategy & position & loss & &PSNR/SSIM & PSNR/SSIM \\
        \midrule
        \ding{56} & - & - & - & 32.04/0.8940 & 28.50/0.7801 \\ 
        \ding{52} &  mid & $\mathcal{L} _{1}$ & 1:1 & 32.11/0.8949 & 28.56/0.7811 \\ 
        \ding{52} &  mid & $\mathcal{L} _{1}$ & 5:1 & 32.11/0.8949& 28.56/0.7811 \\ 
        \ding{52} &  mid & $\mathcal{L} _{2}$ & 1:1& 32.11/0.8949 & 28.56/0.7811 \\ 
        \ding{52} &  end & $\mathcal{L} _{1}$ & 1:1 & \textbf{32.12/0.8951} & \textbf{28.57/0.7813} \\
        \ding{52} &  end & $\mathcal{L} _{1}$ & 5:1 & 32.11/0.8950 & 28.57/0.7813 \\

        \bottomrule
    \end{tabular}}
    \label{tab:distillation}
\end{table}

\noindent\textbf{Teacher model.} When the teacher model has more parameters and richer representation capability, the knowledge it transfers to the student model will be more abundant, leading to a more significant performance improvement of the student model on the task. To verify this conclusion, we attempted two teacher models with different parameters. They exhibited a PSNR difference of 0.27dB on the Set5 dataset. However, as shown in Table~\ref{tab:teacher model}, the performance of the student model remained unchanged. This could indicate that the student model's capacity or architecture may not be sufficiently expressive to fully utilize the additional knowledge provided by the larger teacher model. Therefore, finding the balance point between the performance of the teacher model and the student model is a worthwhile exploration.

\begin{table}[h]
    \centering
    \caption{Design of the teacher model. PSNR is the result of testing on Set5. Params is the parameter of teacher model, and the parameter of student model is fixed.}
    \resizebox{0.9\linewidth}{!}{
    \begin{tabular}{lcccccc}
        \toprule
        \multirow{2}{*}{Method} & \multirow{2}{*}{Params[M]}  & Teacher model & Student model\\
         & & PSNR/SSIM & PSNR/SSIM \\
        \midrule
        DVMSR & -  & - & 32.04/0.8940 \\ 
        DVMSR & 4.089  & 32.38/0.8977 & 32.12/0.8950 \\ 
        DVMSR & 7.432  & 32.65/0.9011 & 32.12/0.8950 \\
        \bottomrule
    \end{tabular}}
    \label{tab:teacher model}
\end{table}

\begin{figure}[h]
    \centering
  \includegraphics[width=0.8\linewidth]{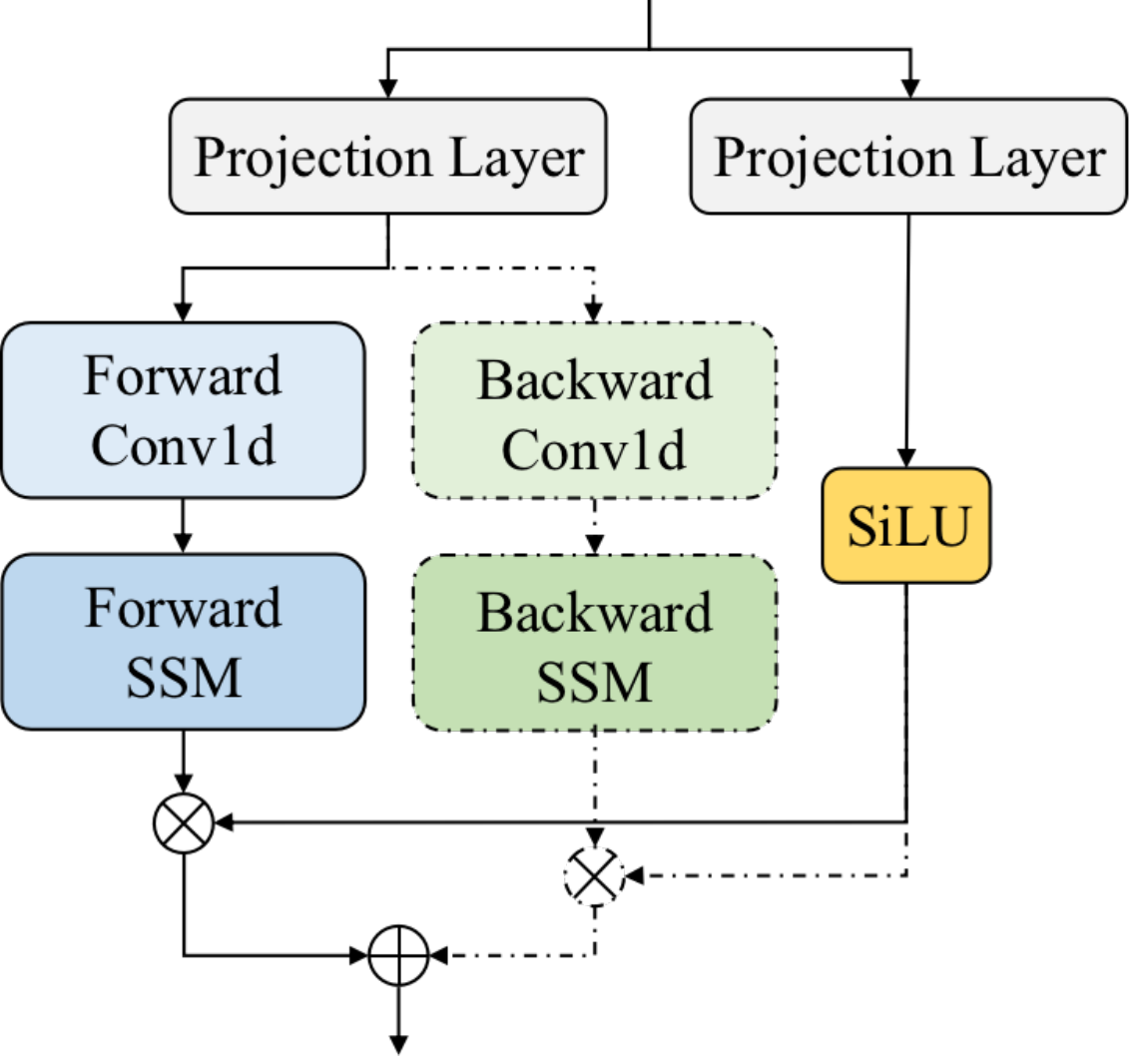}  
  \caption{Unidirectional SSM or bidirectional SSM in ViMM.}
  \label{fig:bivim}
\end{figure}

\subsubsection{Unidirectional v.s. Bidirectional SSM}

To investigate the effectiveness of bidirectional SSM in ESR, we evaluate its performance in ESR based on several aspects: PSNR, Time, Params, FLOPS, Activations, and Memory. The architecture of unidirectional SSM and bidirectional SSM are presented in Figure~\ref{fig:bivim}. As shown in Table~\ref{tab:bissm}, compared to Unidirectional SSM, the improvement of bidirectional SSM in PSNR is limited (increased by 0.01dB), while the inference time has increased by 0.039s. This cost is significant. Therefore, Unidirectional SSM is more suitable for the ESR task.

\subsubsection{NTIRE 2024 Challenge on Efficient SR}

We actively participate in the NTIRE 2024 Efficient Super-Resolution Challenge~\cite{2024ntire}.  The model structure and training strategy are slightly different from the above. This competition aims to procure solutions that excel in overall performance metrics, encompassing inference runtime, FLOPS, and parameter optimization on the NVIDIA GeForce RTX 3090 GPU. This challenge also requires the maintenance or enhancement of threshold PSNR results, underscoring the importance of efficiency without compromising on image quality benchmarks.

During the teacher learning phase, we train the teacher network using the DIV2K dataset with a resolution of 2K. Our teacher architecture consists of 6 RSSB (Residual Scaling and Shifting Block) and 2 ViMM (Vision Mamba Modules), each configured with 180 channels.

In the subsequent distillation training phase, we amalgamated data from both the DIV2K and LSDIR datasets to train the student network. This student model comprises 2 RSSB and 2 ViMM blocks, tailored with 60 channels to maintain computational efficiency while preserving performance standards. Notably, the teacher network remains unchanged.

We employ DIV2K~\cite{div2k} and LSDIR~\cite{lsdir} to construct the training dataset. 
The High-Resolution (HR) images are cropped to $256\times 256$ patches for the training procedure. 

During network optimization, we employ the $\mathcal{L}_{1}$ loss function in conjunction with the Adam optimizer, a widely adopted optimization algorithm in deep learning tasks. Our optimization regimen commenced with an initial learning rate of $2\times 10^{-4}$, evolving through a multi-step learning rate strategy. Specifically, the learning rate halved at key iterations: 250000, 400000, 450000, and 475000, respectively, throughout the 500k total iterations. This adaptive learning rate scheme enhances model convergence and stability over the training period, crucial for achieving superior performance.

Through extensive experiments, we refine our model's architecture and training process, aiming for excellence in both efficiency and performance, as evidenced by our results in Table~\ref{tab:ntire2024}. Our approach employs a novel architecture that differs from both CNN and transformer, providing a reference for the development of mamba in Efficient Super-Reslution.
\begin{table}[h]
    \centering
    \caption{NTIRE 2024 ESR Challenge results.}
    \resizebox{1\linewidth}{!}{
    \begin{tabular}{lcccccc}
        \toprule
        \multirow{2}{*}{Model} & Val PSNR & Test PSNR & Val Time & Test Time & FLOPS & Params  \\
              & (dB) & (dB) & (ms) & (ms) & (G) & (M)  \\
        \midrule
        RLFN\_baseline & 26.96 & 27.07 & 14.348 & 9.194 & 19.67& 0.317  \\
        DVMSR & 26.93 & 27.04 & 40.75 & 34.634 & 20.17 & 0.424  \\
        \bottomrule
    \end{tabular}}
    \label{tab:ntire2024}
\end{table}

\section{Conclusion}
In this paper, we propose DVMSR, a novel lightweight Image SR network that incorporates Vision Mamba and a distillation strategy. It consists of three main modules: feature extraction convolution, multiple stacked Residual State Space Blocks (RSSBs), and a reconstruction module.  In particular, we use a stack of residual state space blocks (RSSB) for deep feature extraction, and each RSSB is composed of Vision Mamba Moudles, a convolution layer and a residual connection. Specifically, we leverage the larger teacher model as additional supervision, which effectively enhances the performance of the student model. DVMSR demonstrates the potential for efficient and long-range dependency modeling in SR tasks, but our work merely offers a preliminary insight. We still need further to explore the potential of Mamba in ESR tasks.
{
    \small
    \bibliographystyle{ieeenat_fullname}
    \bibliography{main}
}


\end{document}